\documentclass[10pt]{article}
\pdfoutput=1
\usepackage{graphicx} 
\usepackage{amsmath} 
\usepackage{amsfonts}
\usepackage{amssymb}
\usepackage{subfigure}
\usepackage{multirow}

\title{ Revealing the Intrinsic Magnetism of Non-Magnetic Glasses  } 

\author{ Giancarlo Jug$^{a,b,c}$ and Sandro Recchia$^{a}$ \\ \\

$^a$Dipartimento di Scienza ed Alta Tecnologia \\ 
Universit\`a dell'Insubria, Via Valleggio 11, 22100 Como (Italy) \\
$^b$INFN -- Sezione di Pavia, Italy \\
$^c$ To.Sca.Lab, Dipartimento di Scienza ed Alta Tecnologia, Universit\`a 
dell'Insubria \\

}

\date{\today} 

\begin{document}

\maketitle

\begin{abstract}
{\bf
Ordinary multi-component silicate glasses belong to a class of amorphous insulators 
normally displaying no special form of magnetism, save for the Larmor dominant
diamagnetism from the constituent atoms' core electrons and the extrinsic Langevin 
paramagnetism due to the ubiquitous Fe-group dilute paramagnetic impurities.  Here 
we show that the macroscopic magnetisation of three case-study glass types 
measured in a SQUID-magnetometer cannot be explained solely by means of the 
Larmor-Langevin contributions. In particular, we reveal a novel {\em intrinsic} 
contribution to the bulk magnetisation due to the amorphous structure itself, a 
contribution that is peculiar both in its temperature and magnetic-field dependence 
and represents the first true magnetic effect in nominally non-magnetic glasses. 
The only theoretical interpretation we know of for such an effect and which can 
consistently explain the experimental data demands the re-thinking of the atomic 
organisation of glasses at the nanometric scale.  
}
\end{abstract}
\newpage


{\bf A. Introduction.} Glasses are ubiquitus substances known to mankind since 
pre-history and yet still presenting many unsolved mysteries to the inquisitive 
scientific mind. What is the precise nature of the atomic structure of glasses, 
especially at medium-range (MR) scales and determining the physical properties of 
these non-crystalline solids? Are glasses just dynamically arrested liquids or rather 
a peculiar new type of solids? What degrees of freedom truly control the onset of 
vetrification, i.e. the glass ``transition'', and convesely: how best to describe the 
melting of a glass? How can we describe theoretically their physical properties on 
the basis of a universal, composition-independent modeling framework that would 
allow for the theoretical control and engineering of the desired glass properties? 
All of these questions and the quest for new technically exploitable or interesting 
physical effects make glasses a very much active and attractive field of research 
in condensed matter physics. 

Where the first of the above questions is concerned -- MR structure -- several 
scenarious have been proposed over the last hundred years 
\cite{Leb1921,Ran1930,Zac1932,War1934,Hag1935,PK1990,Gas1998,Bak1994,
Bak2013,Wri2014,Jug2018}. 
A much acclaimed approach considers (so-called) network glasses as continuously- 
and homogeneously-disordered media at the local (atomic), MR and long-range 
lenght scale: it is the so-called continuous-random-network (CRN) model of 
Zachariasen and Warren \cite{Zac1932,War1934}. However, other scientists have 
suggested to further elaborate upon the CRN approach in favour of heterogeneous 
mesoscopic scenarios \cite{Leb1921,Ran1930,Hag1935,PK1990,Gas1998,Wri2014}. 
Moreover, scenarios have been proposed where the dynamical heterogeneities 
(DH) known to characterise the supercooled liquid state \cite{Edi2000} continue to 
exist (albeit somewhat more statically) also below $T_g$, the nominal temperature 
of glass formation \cite{Bak1994,Bak2013,Jug2018,JLT2021}.
One of these non-CRN scenarios views glasses as made up of solid-like particle 
regions jammed together somewhat below $T_g$ and presenting liquid-like 
regions in-between \cite{Jug2018}. Remarkably, the liquid-like regions between 
the solid ones may become the sources of collective orbital magnetic moments that 
appear to generate an {\it intrinsic} contribution to the glass sample's magnetisation 
mimicking the known Langevin paramagnetism from isolated impurities. 
In the ordinary silicate glasses the latter are typically from Fe and the Fe-group trace 
elements \cite{Bon2015}. The extra intrinsic magnetic moments issue from 
specialized, magnetic-field sensitive tunneling systems (TS) believed to be 
responsible for a plethora of magnetic effects at low temperatures in glasses 
\cite{Jug2018,JBK2016,SXL2021}. These and the more standard TS (the so-called 
two-level systems (TLS) \cite{Phi1981,Esq1998,SXL2021}) might be part of the 
degrees of freedom sought for in order to investigate the physical properties of 
glasses at higher temperatures too \cite{Jug2019}.

In current strategic terms, as it turns out, a better characterisation of the standard 
TLS (which are parasitic yet ubiquitous defects in glasses) is paramount for 
high-technology developments and crucial to awaited breakthroughs. For example in 
fields like the improved detection of gravitational waves \cite{Zmu2012,Iof2015}, 
the fabrication of coherent qubits for superconductor-based quantum computers 
\cite{Kli2018}, and noise-reduction within quantum-information repeaters (quantum 
memories) in fiber-glass optical transmission \cite{Mac2006,Sag2015,Ran2018}. 
A better understanding of the atomic nature of the TS is therefore also long awaited 
for \cite{SXL2021}.

In this Letter we present what we believe to be an important contribution to 
furthering knowledge in all of the above-mentioned basic and advanced research 
areas by providing the first report of the exotic and genuinely-magnetic effect in 
glasses consistent with the theory briefly outlined above \cite{Jug2018,Bon2015,JBK2016}. 
A completely new physical effect in glasses, which is weak but laden with important 
consequences.

We have investigated experimentally and theoretically three case-study systems 
within the class of multi-component silicate glasses: Edmund Optics' BK7 
(boro-silicate best optical glass, with main composition 
K$_2$O-B$_2$O$_3$-Na$_2$O-SiO$_2$); 
Schott's Duran$^{\texttrademark~}$ (boro-silicate best chemistry glass, main 
composition Al$_2$O$_3$-Na$_2$O-B$_2$O$_3$-SiO$_2$) and Heraeus' 
IP-211-clear$^{\texttrademark~}$ (bario-allumino-silicate coating glass, main 
composition CaO-Al$_2$O$_3$-BaO-SiO$2$, hereafter termed BAS in short). 
However, we have good reasons to believe that our findings and their possible 
explanation (with appropriate revealing conditions and different main microscopic 
players) should be pertinent to all amorphous solids (bulk glasses and amorphous 
films, though not necessarily to the non-atomistic colloidal glasses). 

We have systematically investigated the dependence of $M$ (the macroscopic 
magnetisation of point-like samples) on $T$ (temperature) and $H$ (magnetic field), 
in the range 2 $\le T \le$ 315 K and 0.5 $\le H \le$ 65 kOe by means of a 
SQUID-magnetometer and for the above-mentioned glassy solids. In such insulating 
glasses, the SQUID-characterisation is normally employed to asses the trace-like 
(ppm) content of paramagnetic impurities (iron, typically) from a single $M$ vs. 
$1/T$ chart at fixed $H$. The idea, never challenged thus far (however, see 
\cite{Bon2015}), has been that only Fe$^{3+}$ is present in such systems. Then, 
the best-fit to the data with the textbook form \cite{AM1976} (hereafter already 
extended for more than one paramagnetic species):
\begin{eqnarray}
M(T,H)&=&-\chi_L~H+\sum_{s=Fe^{2+},Fe^{3+},\dots}~ 
n_s~g_s\mu_B J_s~ {\cal B}_{J_s}(z_s), \cr
z_s&\equiv&\frac{g_s\mu_BJ_sH}{k_BT}, \cr
{\cal B}_J(z)&=&\frac{2J+1}{2J}\coth\Big(  \frac{2J+1}{2J} z \Big)
-\frac{1}{2J}\coth\Big(  \frac{1}{2J} z \Big)
\label{LL}
\end{eqnarray}
should yield the minute concentrations $n_s$ of Fe- (and other) ions which are 
presumed to be evenly and individually distributed throughout the sample (no 
clustering). In the above formula, $\chi_L$ is the Larmor diamagnetic susceptibility, 
$J_s$ the effective spin value of species $s$ (Fe$^{3+}$ ($J$=5/2), Fe$^{2+}$ 
($J$=2), but in principle also Cr$^{3+}$ ($J$=3/2), Ti$^{3+}$ ($J$=1/2), etc.) 
and $g_s$ its Land\'e's factor. $\mu_B$ is Bohr's magneton, $k_B$ Boltzmann's 
constant and ${\cal B}_J(z)$ is the usual Brillouin function. Because of the best fit,
typically Fe is believed to be present as Fe$^{3+}$ only \cite{Her2000}. But there 
is more to this point of view.

Representative experimental data are 
shown in Fig. \ref{prelim} for the case of a Duran glass' small shard. More details of 
the investigation in the Supplementary Informations (SI) for this paper. In agreement 
with current beliefs \cite{Her2000}, we take $g$=2.0 for both Fe$^{2+}$ and 
Fe$^{3+}$ paramagnetic species in the Larmor-Langevin (LL) best-fit of the data
for silicates. There are cases in the literature where EPR resonances for $g$=1.9 
and 2.1 are found, e.g. for Ti$^{3+}$ in high concentrations for a boro-silicate glass 
\cite{Dav1985}. We stress, however, that in view of the universality tests below, 
using (unascertained) $g\neq$2 values for the Fe-ions would not change our 
qualitative conclusions even when different $g_s$-values for different paramagnetic 
$s$-species are employed. 

\begin{figure}[h!]
\centering
  \vskip -0mm
  \includegraphics[scale=0.30] {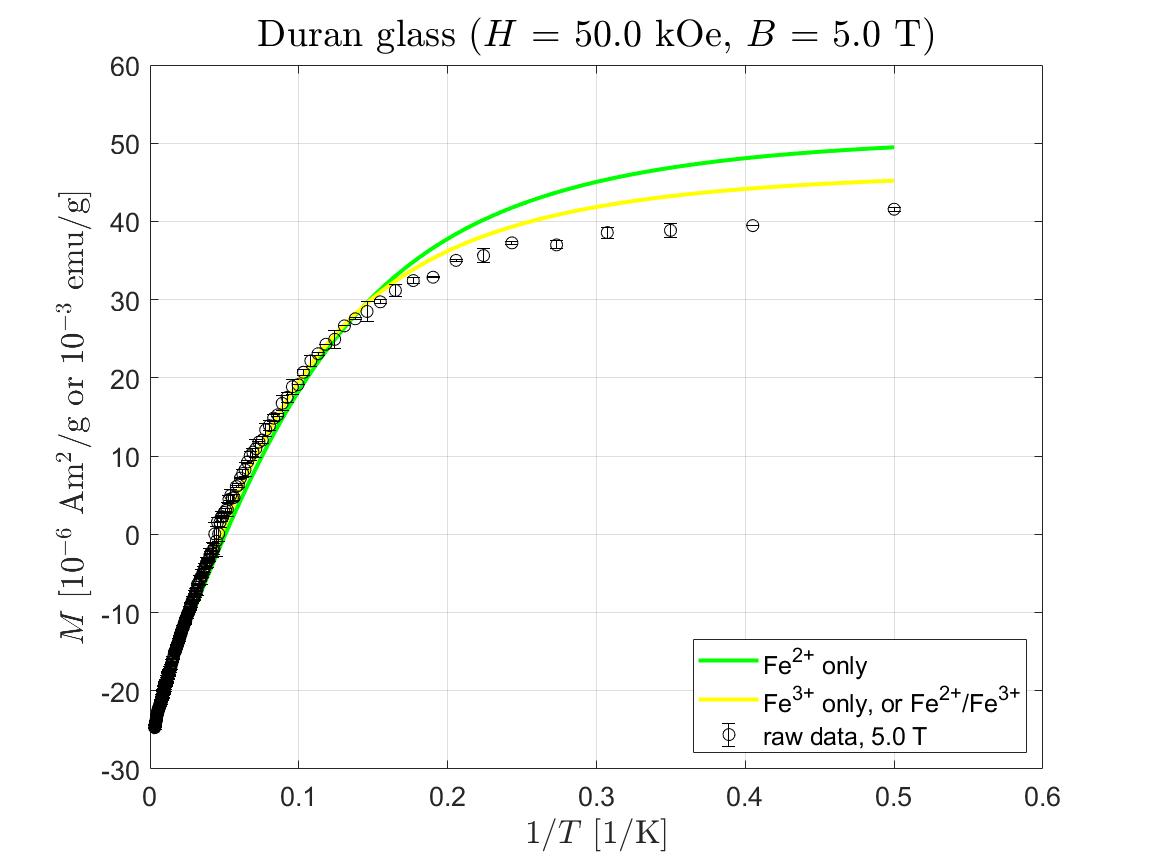}
 \vskip -0mm
\caption{ 
Duran sample's measurement of the magnetisation $M$ at $H$=50.0 kOe 
(5.0 T) in a SQUID-magnetometer. Raw data (black dots, with errorbars) being 
best fitted with Eq. (\ref{LL}) and one paramagnetic species, or two (in which case 
the Fe$^{3+}$-only scenario is always selected). 
}
\label{prelim}
\end{figure}

One would conclude from the above measurement and LL best-fit, Fig. \ref{prelim},
that the sample contains only Fe$^{3+}$ with a concentration
$n$(Fe)=1.539 $\times$ 10$^{18}$ g$^{-1}$ [some 158 ppm] and 
with a Larmor susceptibility $\chi_L$=5.030 $\times$ 10$^{-7}$ emu/gOe (or 
cm$^3$/g). Though the above is one of our worst-case examples, the agreement
between our data and expression (\ref{LL}) is generally unsatisfactory for all studied 
samples, indicating that an extra contribution is missing. We also notice that: 
1) The data agreement with the LL form, Eq. (\ref{LL}), is generally worst at the 
lowest temperatures and fixed $H$, even for the Fe$^{3+}$-only scenario and 
will not improve by allowing other contaminants having $J_s<$5/2 in the best-fit.
Imposing the Fe$^{2+}$-only scenario (thus yielding 2.027 $\times$ 10$^{18}$ 
g$^{-1}$ [some 208 ppm] of Fe) makes the disagreement much worse; 
2) This naively SQUID-extracted Fe-concentration is somewhat in disagreement with 
the result from our own very precise mass-spectrometry (Inductively-Coupled Plasma 
Mass Spectrometry, ICP-MS, hereafter referred simpy as MS) analysis of the sample 
giving $n$(Fe)=(1.469 $\pm$ 0.006) $\times$ 10$^{18}$ g$^{-1}$ only
[meaning 151.0 $\pm$ 0.6 ppm]; 
moreover, a quick textbook-formula \cite{AM1976} estimate of Larmor
susceptibility yields for Duran: $\chi_L \simeq$ 4.03 $\times$ 10$^{-7}$ emu/gOe; 
3) It is very surprising that Fe$^{3+}$ should be considered predominant in the 
Duran shard, because of the faint greenish tinge of the specimen (Figs. 1(a--c) in the 
SI) like in most ordinary clear silicate glass items which indicates rather Fe$^{2+}$ 
predominance \cite{Var1994}.

We have carried out a range of SQUID-runs both with $H$- and $T$-fixed, reaching 
the conclusion that LL-based theory fits of SQUID-data are unreliable for glasses 
without a fuller interpretation (more below). Moreover, as we shall demonstrate in 
this paper, the missing contribution that comes from the glass itself when 
hypothetically cleaned of all paramagnetic impurities is a new and rather fascinating 
phenomenon with important implications for glass science and technology. 

\vfill
{\bf B. Full BK7 glass analysis.} As expected and as shown below, the case of BK7
is that of a very pure and micro-crystal free glass, so we have conducted most of
our study on this system. 
Tables 1 and 2 as well as Fig. 7  in the SI show significant fluctuations in both 
$n$(Fe) and $\chi_L$ by SQUID-scanning in $T$ or $H$ while keeping 
$H$ or $T$ constant, respectively, and employing the LL-fit form, Eq. (\ref{LL}).

\begin{table}[!htbp]
\begin{center}
\begin{tabular}{ |c|c|c|c|c|c|c|c|c|c|c| } 

\hline

Magnetic Field (kOe) & 1.0 & 2.5 & 5.0 & 10.0 & 20.0 & 30.0 & 40.0 & 50.0 & 
 65.0 & MS \\
\hline

BK7 LL-parameters &  &  &  &  &  &  &  &  &  &  \\
\hline

n(Fe$^{3+}$) 10$^{17}$ g$^{-1}$ & 1.888 & 1.849 & 1.845 & 1.849 & 1.867 &
1.888 & 1.906 & 1.914 & 1.931 & 1.657 $\pm$ 0.019 \\ 

$\chi_L$ 10$^{-7}$ emu/gOe & 3.922 & 4.064 & 4.139 & 4.257 & 4.277 & 4.283 &
4.295  & 4.294 & 4.297 & - \\

\hline

n(Fe$^{2+}$) 10$^{17}$ g$^{-1}$ & 2.752 & 2.690 & 2.667 & 2.621 & 2.559 &
2.541 & 2.536 & 2.526 & 2.526 & 1.657 $\pm$ 0.019 \\

$\chi_L$ 10$^{-7}$ emu/gOe & 3.922 & 4.064 & 4.136 & 4.249 & 4.262 & 4.266 &
4.277 & 4.276 & 4.278 & - \\

\hline

\end{tabular}
\caption{ LL-fitting parameters (Eq. (\ref{LL})) extracted from different SQUID runs 
of $M$ vs. $1/T$ at stated $H$-values for a BK7-prism chip. 
The first row is obtained from best fits when both Fe$^{3+}$ and Fe$^{2+}$ 
species are allowed (in practice one always obtains $n$(Fe$^{2+}$)=0); for 
comparison, the results when Fe$^{2+}$-only is allowed are shown in the second 
row. The precise MS value of $n_{MS}$(Fe) is also indicated (errorbar was 
estimated from spectrometer specifications and by dissolving several separate 
BK7-prism chips, including this one). }
 \label{BK7LLH}
\end{center}
\end{table}

\begin{table}[!hbp]
\begin{center}
\begin{tabular}{ |c|c|c|c|c|c|c|} 

\hline

Temperature (K) & 2.0 & 4.5 & 7.5 & 10.0 & 20.0 & MS \\
\hline

BK7 LL-parameters &  &  &  &  &  &  \\
\hline

n(Fe$^{3+}$) 10$^{17}$ g$^{-1}$ & 1.746 & 1.877 & 1.969 & 2.035 & 3.135 &
1.657 $\pm$ 0.019 \\ 

$\chi_L$ 10$^{-7}$ emu/gOe & 4.178 & 4.269 & 4.326 & 4.350 & 4.686 & - \\

\hline

n(Fe$^{2+}$) 10$^{17}$ g$^{-1}$ & 2.414 & 2.905 & 3.358 & 3.681 & 6.206 &
1.657 $\pm$ 0.019 \\

$\chi_L$ 10$^{-7}$ emu/gOe & 4.178 & 4.269 & 4.326 & 4.350 & 4.686 & - \\

\hline

\end{tabular}
\caption{ The same as in Table \ref{BK7LLH}, but as extracted from SQUID-runs of 
$M$ vs. $H$ at stated $T$-values. }
 \label{BK7LLT}
\end{center}
\end{table}

From the naive SQUID-characterisation one would deduce (say, at $H$=65.0 kOe 
(6.5 T)) concentrations $n$(Fe$^{2+}$)=0 and 
$n$(Fe$^{3+}$)=1.931 $\times$ 10$^{17}$ g$^{-1}$ [some 21 ppm] while 
from our precise MS analysis we find
$n_{MS}$(Fe)=(1.657 $\pm$ 0.02) $\times$ 10$^{17}$g$^{-1}$ [only  
17.5 $\pm$ 0.2 ppm]. The disagreement with the (true) MS value of $n$(Fe) is again  
particularly severe when obtained by imposing Fe$^{2+}$-only (majority species, 
as it ought to be \cite{Var1994}), yielding 2.526 $\times$ 10$^{17}$ g$^{-1}$ 
[some 27 ppm] of Fe at 65.0 kOe) and especially for $M$ vs. $H$ SQUID-runs 
(Table \ref{BK7LLT}).  But even for 
Fe$^{3+}$-only and for $M$ vs. $1/T$ the disagreement is remarkable and 
$n$(Fe$^{3+}$) appears to be $H$-dependent (Table \ref{BK7LLH}) or
$T$-dependent (Table \ref{BK7LLT}).

A way to resolve the discrepancy might be to increase the parameter space and 
include other (minority) paramagnetic species. Some of which are indeed present, 
as shown in Table \ref{qual-MS}. 

\begin{table}[!htp]
\begin{center}
\begin{tabular}{ |c|c|c|c|c|c|c|c| } 

\hline

{\bf Fe} & Ti & V & Cr & Mn & Co & Ni & Cu \\

\hline

{\bf 1} & 0.1 & 0 & 0.1 & 0.01 & 0 & 0.01 & 0.01 \\

\hline

\end{tabular}
\caption{ Relative concentration (taking $n$(Fe)=1) of all other Fe-group magnetic
elements present in the same BK7-chip SQUID-characterised in this work and from 
our own qualitative MS analysis. }
 \label{qual-MS}
\end{center}
\end{table}

However, only Ti (with a $J_s$=1/2 if Ti$^{3+}$ (which is doubtful in the silicates)) 
and Cr$^{3+}$ ($J_s$=3/2) are present in noticeable concentrations and their 
presence in Eq. (\ref{LL}) does not improve the best-fits at all, since one obtains 
again that only Fe$^{3+}$ should be present (due to its highest $J_s$ available) and 
all other species with $J_s<$5/2 have $n_s$=0. An improvement would follow were 
other paramagnetic species with $J_s\ge$5/2 also present (like Mn$^{2+}$, 
Co$^{2+}$ ... or the rare-earth metals) but these are basically all absent. 

\begin{figure}[!htp]
\centering
{ \vskip 0cm
   \subfigure[]{\includegraphics[scale=0.30]  {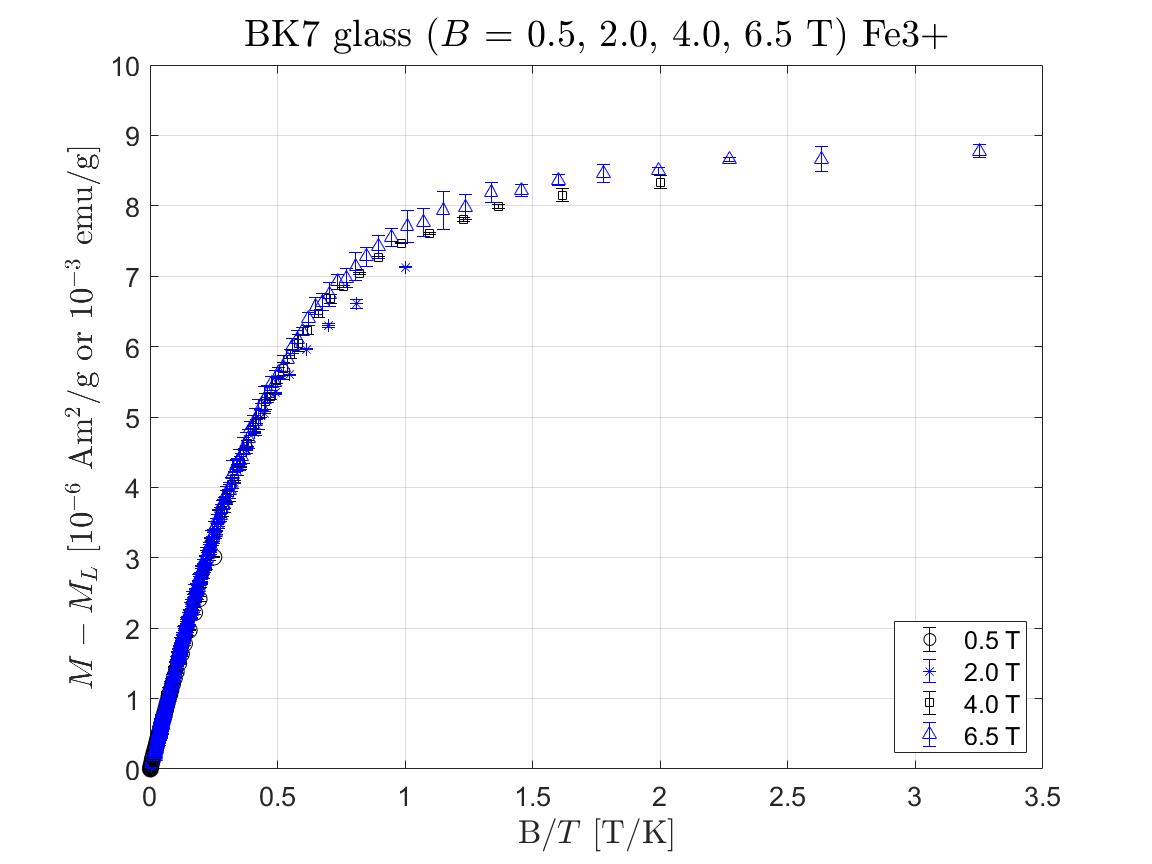} } 
 \vskip -0cm
   \subfigure[]{\includegraphics[scale=0.30] {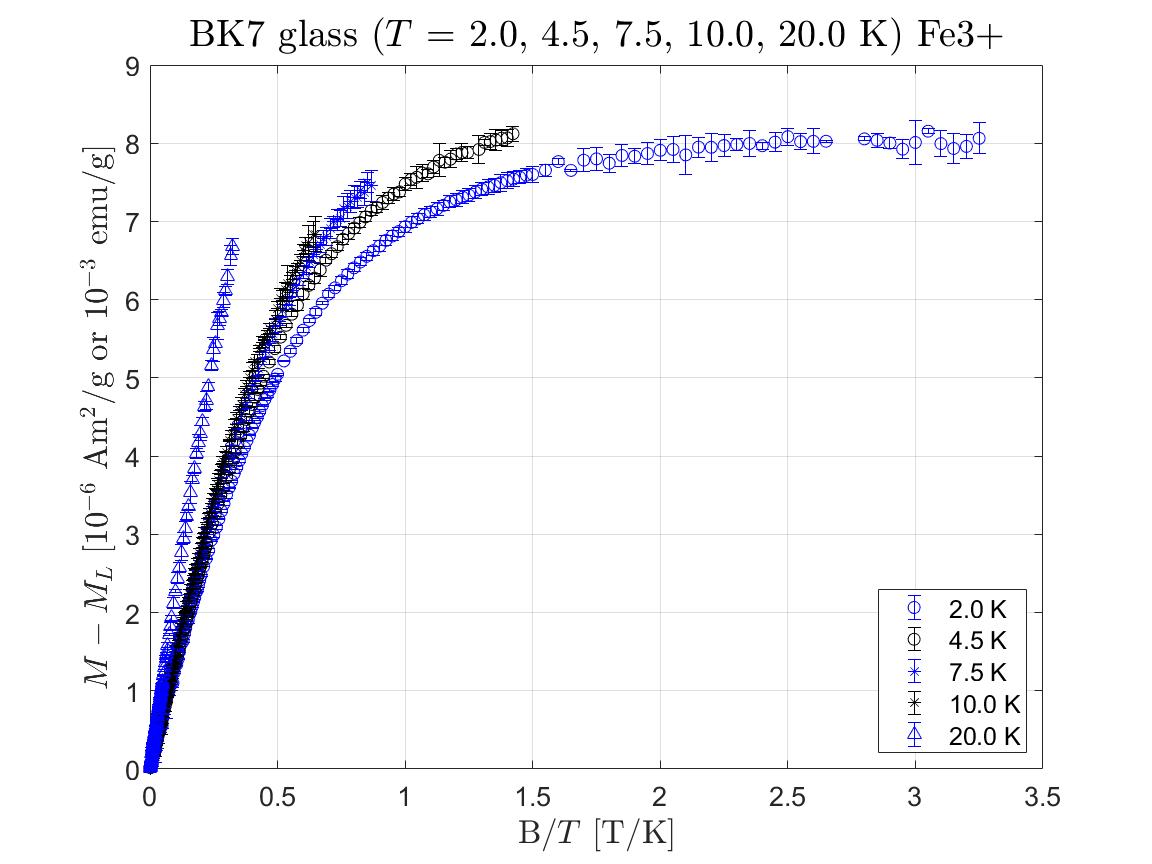} }
}
\caption{ Universality test for our SQUID-run measurements on BK7: (a) with fixed 
$H$ as stated (in Tesla) and (b) with fixed $T$ as stated. For clarity, in
panel (a) not all of our fixed-$H$ runs are reported. Fe3+ means that the 
subtracted Larmor values refer to the best-fit of raw data with the extended LL 
expression, Eq. (\ref{LL}), which always singles out the Fe$^{3+}$-only scenario.
As always, here $B=\mu_0 H$.
}
\label{univ}
\end{figure}

To get to the root of the discrepancy, we have re-plotted all 
available SQUID-runs data as $M-M_L$ ($M_L=-\chi_L H$) vs. $H/T$ which according 
to the generalised LL-expression in Eq. (\ref{LL}) should yield a universal curve. 
This, however, is not the case as seen in Fig.s \ref{univ}(a),(b) and particularly for 
fixed $T$ data.

Clearly there is a contribution to $M(T,H)$ from the sample that is not accounted for 
by the LL-expression Eq. (\ref{LL}), regardless of how many paramagnetic species 
get included. The unaccounted difference $M(T,H)-M_{LL}(T,H)$ is {\em intrinsic} to 
the glass itself and not vanishing should all paramagnetic impurities be removed from 
the sample. The unaccounted term, $M_{intr}(T,H)$, is clearly not a function of 
$H/T$ alone, but has pieces that depend separately on $T$ and $H$. At small $H/T$ 
this contribution must account for the disagreement between SQUID's LL-assessed 
$n$(Fe) values and the MS's $n$(Fe) values (and so $M_{intr}(T,H)$ mimicks the 
Langevin paramagnetic behaviour in this limit \cite{Bon2015}), but at lower $T$ or 
higher $H$ something new must emerge. We remark that clustering of Fe$^{3+}$ 
ions, e.g. in O=Fe-O-Fe=O formation, would just ``segregate'' some Fe in the 
glassy net and lead to lower, not higher, SQUID-ascertained $n$(Fe) values.

\begin{figure}[!htp]
\centering
{ \vskip 0cm
   \subfigure[]{\includegraphics[scale=0.33]  {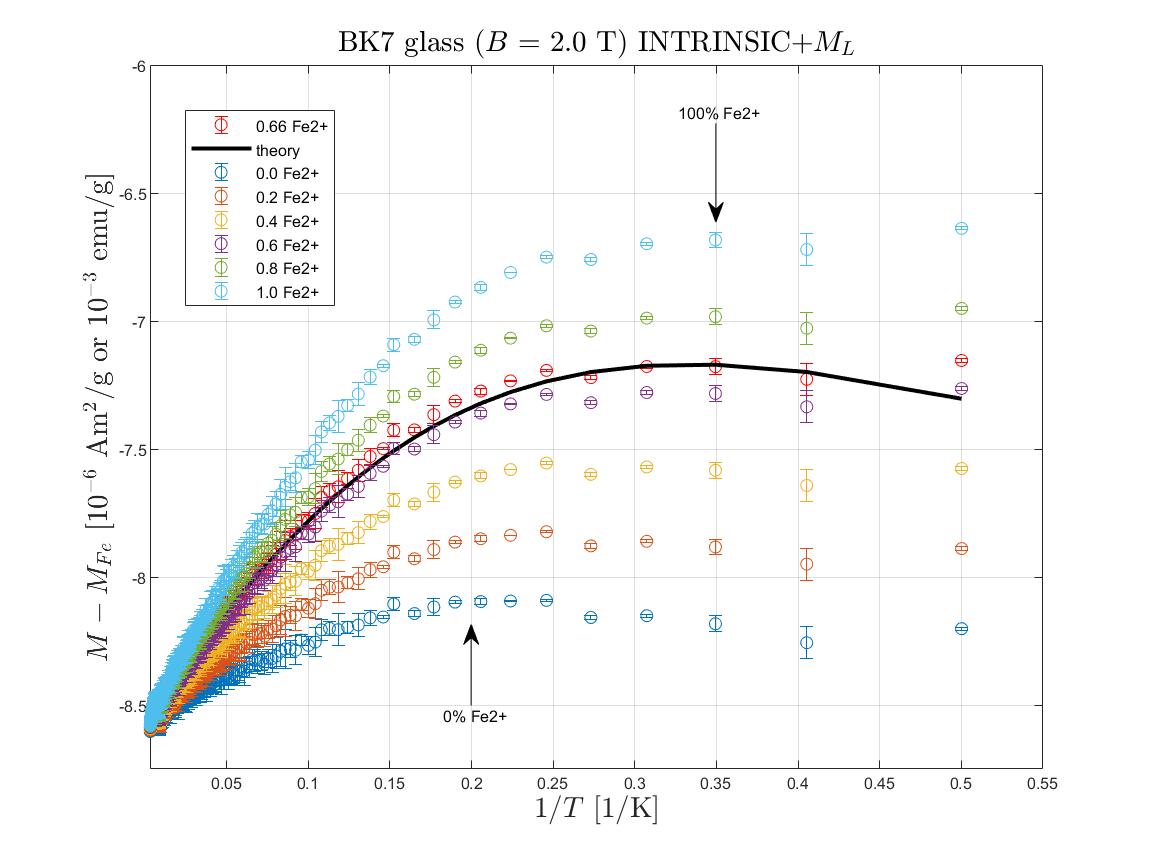} } 
 \vskip -0cm
   \subfigure[]{\includegraphics[scale=0.33] {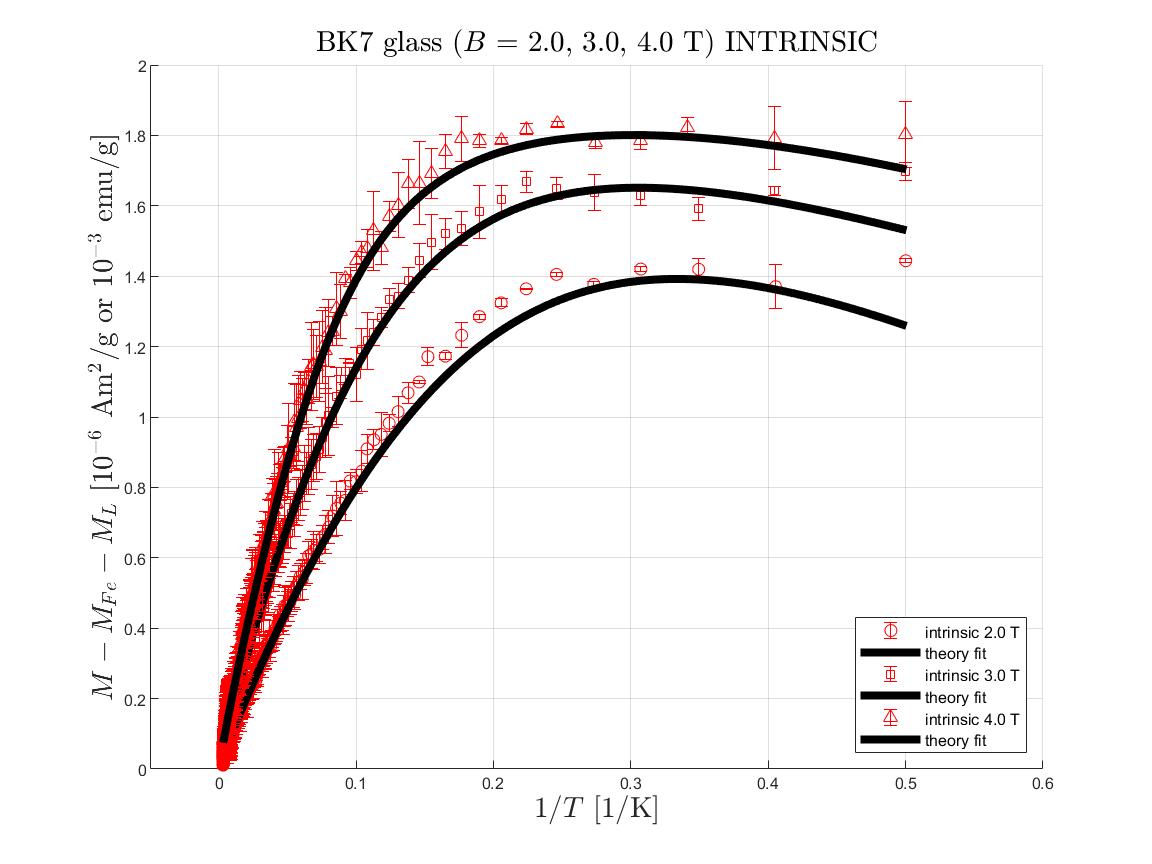} }
}
\caption{ (a) Re-plot of raw data after subtraction of the Langevin contribution from 
$x~n_{MS}$(Fe)  Fe$^{2+}$-ions and $(1-x)~n_{MS}$(Fe) Fe$^{3+}$-ions with 
$n_{MS}$(Fe)=1.66 $\times$ 10$^{17}$ g$^{-1}$ (the MS value, see text); from 
top to bottom $x=x_2$=1.0, 0.8, 0.6, 0.4, 0.2, 0.0; the value $x$=0.66 is selected 
by theory (see below) and the theory fit (full curve) is also drawn. (b) At fixed value 
$x=x_2$=0.66 selected by theory, the re-plotted $M_{intr}$ data points (obtained
by subtraction from raw data) shown for various $B$-values ($B=\mu_0 H$) and 
compared to theory fitting curves. 
}
\label{intrH}
\end{figure}

\begin{figure}[!htp]
\centering
{ \vskip -0cm 
   \subfigure[]{\includegraphics[scale=0.35]  {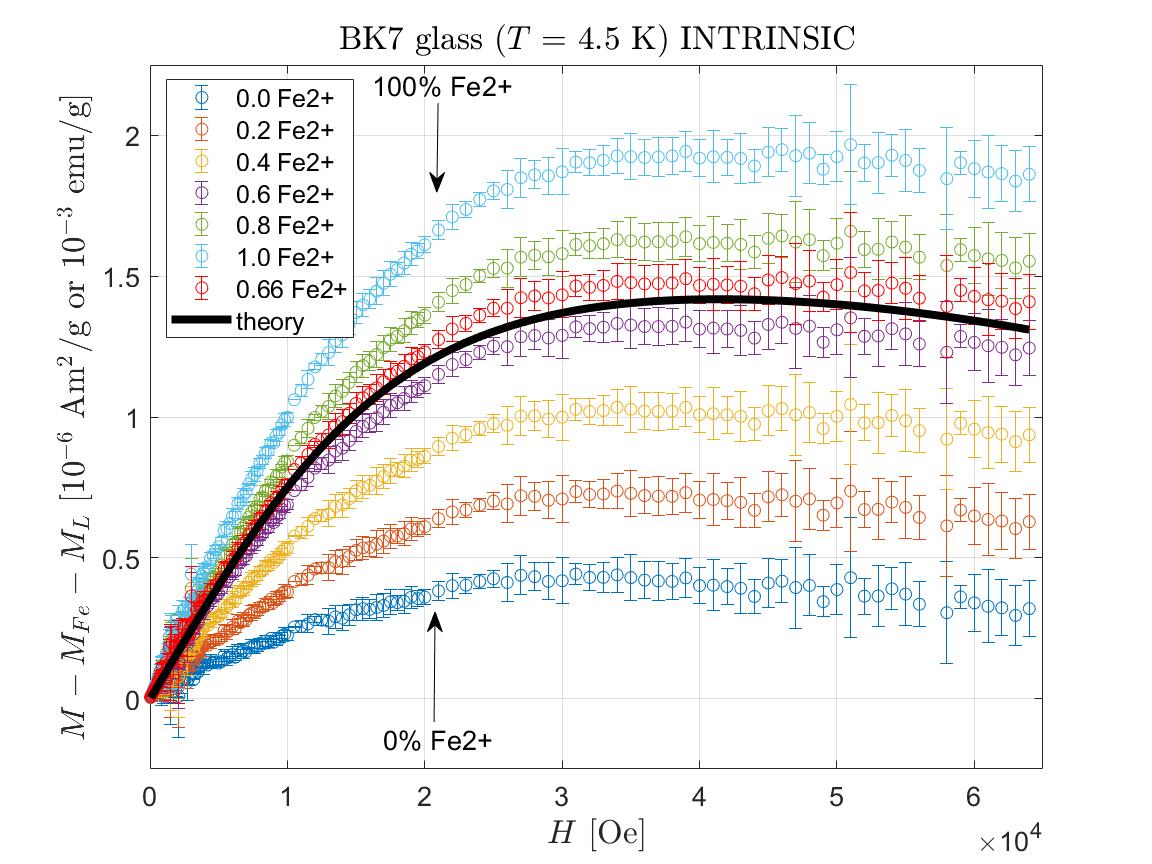}} 
 \vskip -0cm
   \subfigure[]{\includegraphics[scale=0.35] {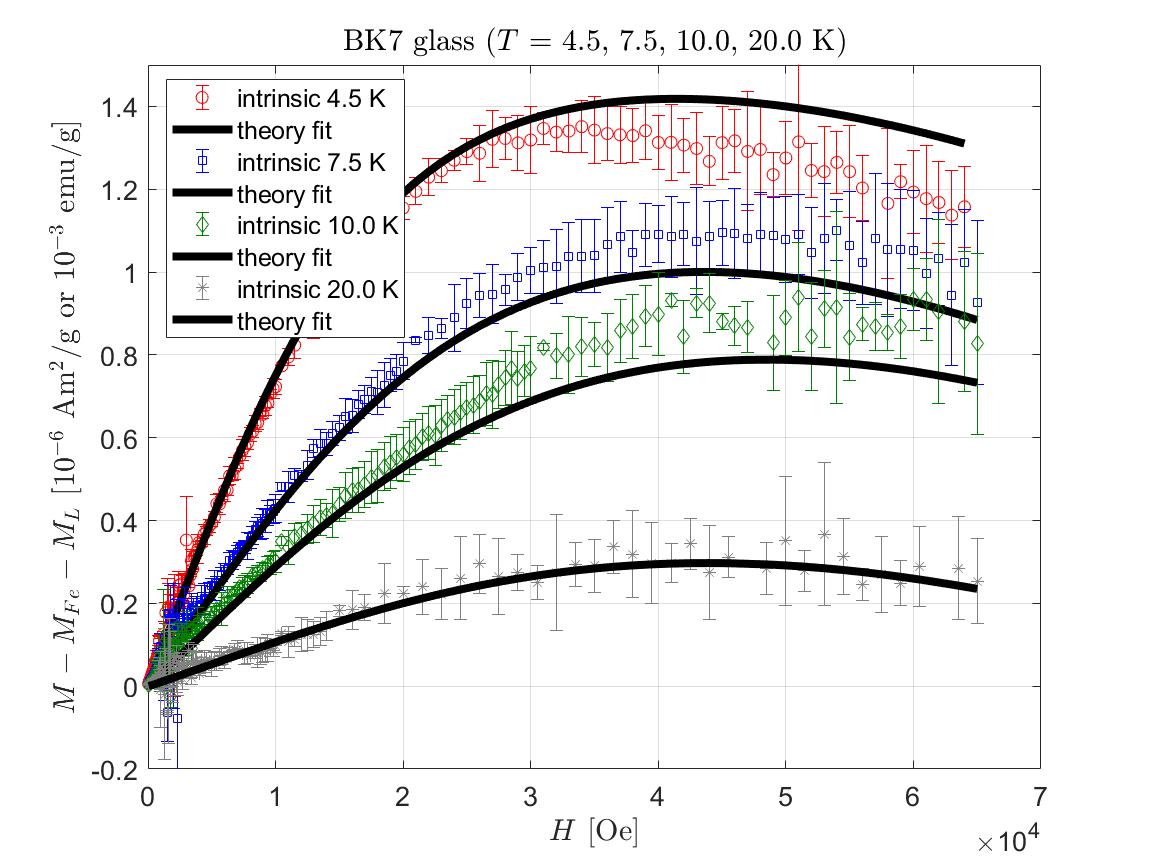} }
}
\caption{ (a) As for Fig. \ref{intrH}(a), but for $M_{intr}$ at fixed $T$=4.5 K.  
(b) As for Fig. \ref{intrH}(b), but  for various $T$-values and compared to theory 
fitting curves. On the horizontal axis $H$ in Oe ($\times$ 10$^4$) or $B$ in T 
($B=\mu_0 H$, as always). 
}
\label{intrT}
\end{figure}

In order to reveal what $M_{intr}(T,H)$ may look like, we simply subtract the 
Langevin contribution (or full LL-contribution to improve clarity) with the MS known 
$n$(Fe) concentration ((1.657 $\pm$ 0.019) $\times$ 10$^{17}$ g$^{-1}$ in this 
case) and best-fit determined $\chi_L$ susceptibility values. 
The relative \% fractions $x_2$ of Fe$^{2+}$ and $x_3$ of Fe$^{3+}$ 
($x_2$+$x_3$=1) are difficult to ascertain at such low concentrations. EPR 
requires interpretation and knowledge of the precise location of the two species in 
the glassy net (which is lacking).  M\"ossbauer spectroscopy also does not help at 
such low Fe-concentrations. Thus we resort to plotting $M_{intr}(T,H)$-data by 
subtracting variable fractions $x=x_2$ of Fe$^{2+}$ and $1-x=x_3$ of 
Fe$^{3+}$ Langevin-contributions. In Fig. \ref{intrH}(a) we present one case of 
what we observe at $H$=20.0 kOe (2.0 T) for BK7 as a function of $x$ (more cases 
being presented in the SI for the other glasses). Also, in Fig. \ref{intrT}(a) we show 
the case $T$=4.5 K as a function of $x$ (more cases in the SI).

We observe a peculiar, exotic behaviour of the ensuing intrinsic magnetisation: both 
as a function of $1/T$ and as a function of $H$,~ $M_{intr}$ displays a broad peak 
instead of saturating to a constant at low temperatures and, respectively, at high 
magnetic fields as ordinary spin- and also orbital-paramagnetism should do. As 
Figs. \ref{intrH}(a) and \ref{intrT}(a) clearly show, this unusual broad peak is 
always present and is independent of $x=x_2$, so this first conclusion comes from 
no theoretical input at all. To fit the data carefully -- understanding the physical origin 
of the broad peaks, how they move by changing $H$ and $T$, respectively, and 
explaining the Fe-concentration mismatch -- a new theory approach is however 
required.

\vfill
{\bf C. Looking through other glasses.} As stated in the Introduction, we have also 
investigated samples of Duran and BAS glasses which, however, turned out to have 
much higher Fe-contamination. Nevertheless, our conclusions remain the same as 
for BK7 and receive much increased confidence. 

\begin{figure}[!htp]
\centering
{ \vskip -0cm 
   \subfigure[]{\includegraphics[scale=0.60]  {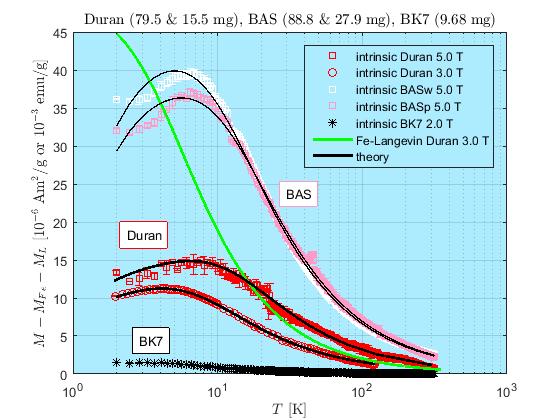}} 
 \vskip -0cm
   \subfigure[]{\includegraphics[scale=0.60] {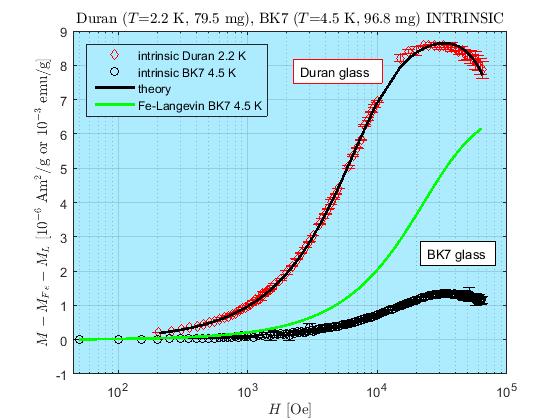} }
}
\caption{ (a) A few ``experimental'' curves (with theory fits) for $M_{intr}$ for the 
two other glasses, Duran and BAS, as a function of $T$ and for the $x=x_2$ 
Fe$^{2+}$-fraction selected by the theory fit (BK7 also shown for comparison): 
$x_{\rm BK7}$=0.66, $x_{\rm Duran}$=0.83 and $x_{\rm BAS}$=0.63 (both -p 
and -w type). The green curve is the Langevin Fe-contribution for Duran at 3.0 T. 
(b) ``Experimental'' curve (with theory fit) for $M_{intr}$ for Duran, as a function of 
$H$ (BK7 also shown for comparison): $x_{\rm BK7}$ and $x_{\rm Duran}$ as in 
(a).  More cases and theory comparison in the SI. The green line is the Fe-Langevin 
for BK7 at 4.5 K. Sample masses serve to identify different samples.} 
\label{otherglasses}
\end{figure}

As is shown in the SI, the same lack of universality for $M(T,H)$ data sets as 
functions of $H/T$ re-presents itself, particularly severe for Duran. Dissolving 
several Duran glass shards from the same item, we were able to determine, from 
MS, $n_{MS}$(Fe)=(1.469 $\pm$ 0.006) $\times$ 10$^{18}$ g$^{-1}$ 
[or 151.0 $\pm$ 0.6 ppm] which we used to chart out some of the $M_{intr}(T,H)$ 
glass-intrinsic magnetisation behaviour (Fig. \ref{otherglasses} and in the SI). In spite 
of the much higher Fe-content, $M_{intr}(T,H)$ resulted to be also much more 
intense and the broad peaks more pronounced than in BK7. 
This is exemplified in Fig.s \ref{otherglasses}(a) and \ref{otherglasses}(b) where, 
for comparison, a case for BK7 and the cases for BAS-p and BAS-w, see below, are 
also reported. Again, the broad peaks in the $T$- and $H$-dependences of 
$M_{intr}$ are confirmed and, as shown in the SI, are also qualitatively independent 
of $x$ (Fe$^{2+}$-fraction in $n$(Fe)).
 
We have also sintered and vitrified two separate samples, in the form of small 
pellets, of Heraeus's IP-211-clear (a type of BAS) paste following the manufacturer's 
instructions. One of the samples, which we term BAS-p, presented a glassy 
appearance and residual pink coloration (Fig. 1(a) in the SI). The other -- 
the sol-gel paste dried much longer and calcinated at 1000 C before vitrification --
presented a perlaceous appearance and residual white coloration and we term this 
BAS-w. The coloration is an indication of the average size of the fused multi-silicate 
micro-spheres present within the sol-gel original paste (of a dark-violet colour). 
Hence, we retain that the BAS-p sample had a larger -- and less dispersed --
average-size in the final micro/mesoscopic clustering than the BAS-w sample (see 
theory below). 
 
The most interesting results (Fig. \ref{otherglasses}) appear to be for the BAS 
samples, where the intensity of the 
intrinsic glass-magnetism appears to increase at the lowest $T$ with heat treatment 
(in going from BAS-p to BAS-w). As for the Fe-concentration, from mass 
spectroscopy we were able to determine a staggering 
$n_{MS}$(Fe)=(7.165 $\pm$ 0.032) $\times$ 10$^{18}$ g$^{-1}$ [958.1 $\pm$ 4.3 
ppm] from a series of calcinated BAS-paste flakes. This again contrasts with a naive
SQUID-magnetisation LL-ascertained value (e.g. at $H$=50.0 kOe) 
$n$(Fe$^{3+}$)=7.295 $\times$ 10$^{18}$ g$^{-1}$ [some 976 ppm] (or 
$n$(Fe$^{2+}$)=9.634 $\times$ 10$^{18}$ g$^{-1}$ [some 1288 ppm] in both 
cases with a bad fit at the lower temperatures). 

We characterised $M_{intr}(T,H)$ using the same subtraction procedure as for BK7.
Despite the much increased Fe-contamination, for both Duran and BAS systems the 
intrinsic contribution remains clearly ascertained and for BAS it is the strongest for 
the three systems investigated. All three systems display the possible presence of 
oscillations in $M_{intr}(T,H)$ vs. $T$ at the lowest temperatures investigated 
($T<$ 3 K, see the SI).

\begin{figure}[h!]
\centering
  \vskip -0mm
  \includegraphics[scale=0.60] {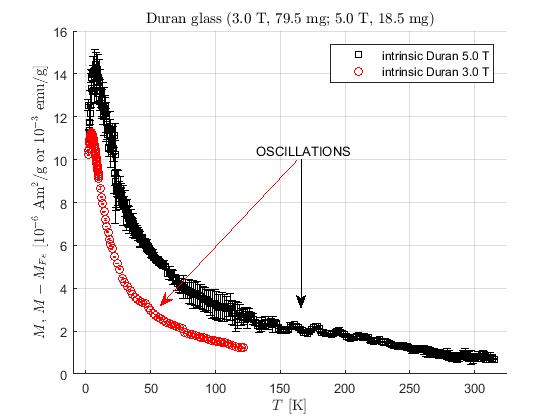}
 \vskip -0mm
\caption{ 
Duran samples' raw-data deduced intrinsic magnetisation (after subtraction of 
electronic-Larmor and Fe-Langevin contributions) as a function of temperature for 
$B$=3.0 T (red symbols) and for 5.0 T (black). Distinct oscillations in the 
intermediate temperature range are clearly visible, especially for the 5.0 T case, 
and well above the instrumental errorbar range). They are also observed for the
BAS-p sample (see SI). 
}
\label{oscill}
\end{figure}

\vfill
{\bf D. Intrinsic magnetisation temperature oscillations.} To conclude the survey of
our novel observations, we report that by re-plotting the intrinsic magnetisation 
$M_{intr}(H,T)$ as a function of temperature $T$ in a linear scale some distinct 
oscillations for fixed $H$ as a function of $T$ in the intermediate range are observed.
These cannot be mistaken with spurious instrumental effects and are clearly visible
above the SQUID intrumental errorbars. Moreover, they are present for all three 
glassy systems investigated and in different conditions. In Fig. \ref{oscill} we report 
the clearest cases, those for borosilicate Duran at $B$=3.0 T and especially at 5.0 T. 
As discussed in the SI, we believe the origin of these oscillations lies in the dynamics 
of the non-spherical solid-like cells making up the amorphous solid and in the 
response to the cells' dynamics of the charged tunneling (magnetic-field coupled) 
fluid particles in the cell-cell interspace  ``voids''. Interestingly, these oscillations may
well be related to the  ``Boson peak'' phenomenology \cite{SXL2021} and have the 
same basic explanation.

\vfill
{\bf E. Proposed theoretical explanation.} There is no standard explanation that we 
know of, for all these observations in the three investigated glass types. Clustering 
of Fe-ions is not an explanation, for $n$(Fe) as determined from the naive LL-fitting 
of SQUID-data is always greater than that determined by the MS. Kramers 
doubling of the $J$=5/2 Fe$^{3+}$-spin spectrum would not help, for it would 
result in a linear dependence of the energy levels and splittings on $H$. Quenching 
of the orbital moments for Fe-ions by the local-``crystal'' field is believed to be 
complete in the multi-silicates anyway \cite{Her2000}, a reason why $g$=2 is 
expected to be correct here, for both Fe$^{2+}$ and for Fe$^{3+}$ \cite{Kit2005}.

The observed features of the intrinsic magnetisation $M_{intr}(T,H)$ are however
quantitatively consistent with the results one obtains from the theoretical scenario 
of intermediate-range glass structure \cite{Jug2018,JBK2016,Jug2004,Jug2009,JPB2014}
briefly outlined in the Introduction. The theory was proposed by one of us (GJ) for 
the explanation of the puzzling  magnetic effects in dielectric (and organic) glasses
\cite{Jug2004}. There is already extensive literature on this theory, which for the 
magnetisation is the continuation of previous work \cite{Bon2015} concerning the 
mismatch in $n$(Fe) as deduced from published \cite{Her2000,Sie2001} SQUID-data 
for BK7-Duran-BAS samples (not our own) and from published \cite{Sie2001} 
heat-capacity data in the 300 mK - 4 K range for the Duran and BAS glasses. The
present work thus sets that theory to a new test, with a much wider range of 
experimental data for $M=M(T,H)$ now available. Once more, this theory carefully
resolves the Fe-concentration mismatch (now between MS and naive LL SQUID-determined 
values). We indeed obtain, fitting our own SQUID-data with 
$M=M_{LL}+M_{intr}^{tunn}$ (the latter contribution being provided by theory 
discussed below and in the SI): 
for BK7
$n$(Fe)=(1.688 $\pm$ 0.004) $\times$ 10$^{17}$ g$^{-1}$ [17.8 $\pm$ 0.1 ppm];
for Duran
$n$(Fe)=(1.471 $\pm$ 0.006) $\times$ 10$^{18}$ g$^{-1}$ [151.1 $\pm$ 0.6 ppm];
for BAS
$n$(Fe)=(7.127 $\pm$ 0.007) $\times$ 10$^{18}$ g$^{-1}$ [953.1 $\pm$ 0.9 ppm]. 
The agreement with the data obtained from the MS analysis is thus now rather 
satisfactory. As is shown in the SI, the tunneling parameters are roughly the same 
as those obtained for the BK7-Duran-BAS 2000 samples \cite{Bon2015}, the major 
difference being the Fe-concentrations which are higher for our own samples. As for 
the strength of the tunneling intrinsic contribution, we obtain
$n_{tunn}/n_{MS}$(Fe) concentration ratios for the three systems we investigated:
BK7 (1.348$\times$10$^{16}$/1.657$\times$10$^{17}$)=0.081 (8.1 \%);
Duran (4.519$\times$10$^{16}$/1.469$\times$10$^{18}$)=0.031  (3.1 \%);
BAS (2.323$\times$10$^{17}$/7.165$\times$10$^{18}$)=0.032 (3.2 \%).
This justifies our choice to concentrate our study on BK7: though the intrinsic 
magnetisation is the weakest, in comparison the Fe-concentration is the lowest and 
the ratio of intrinsic/Langevin contributions the highest for this system.
Moreover, as shown in the graphs in this article, the tunneling theory explains 
rather well most of the new features of the intrinsic magnetisation of glasses.

\vskip 0.5cm

\begin{figure}[!hbp]
\centering
{ \vskip -1cm 
   \subfigure[]{\includegraphics[scale=0.16]  {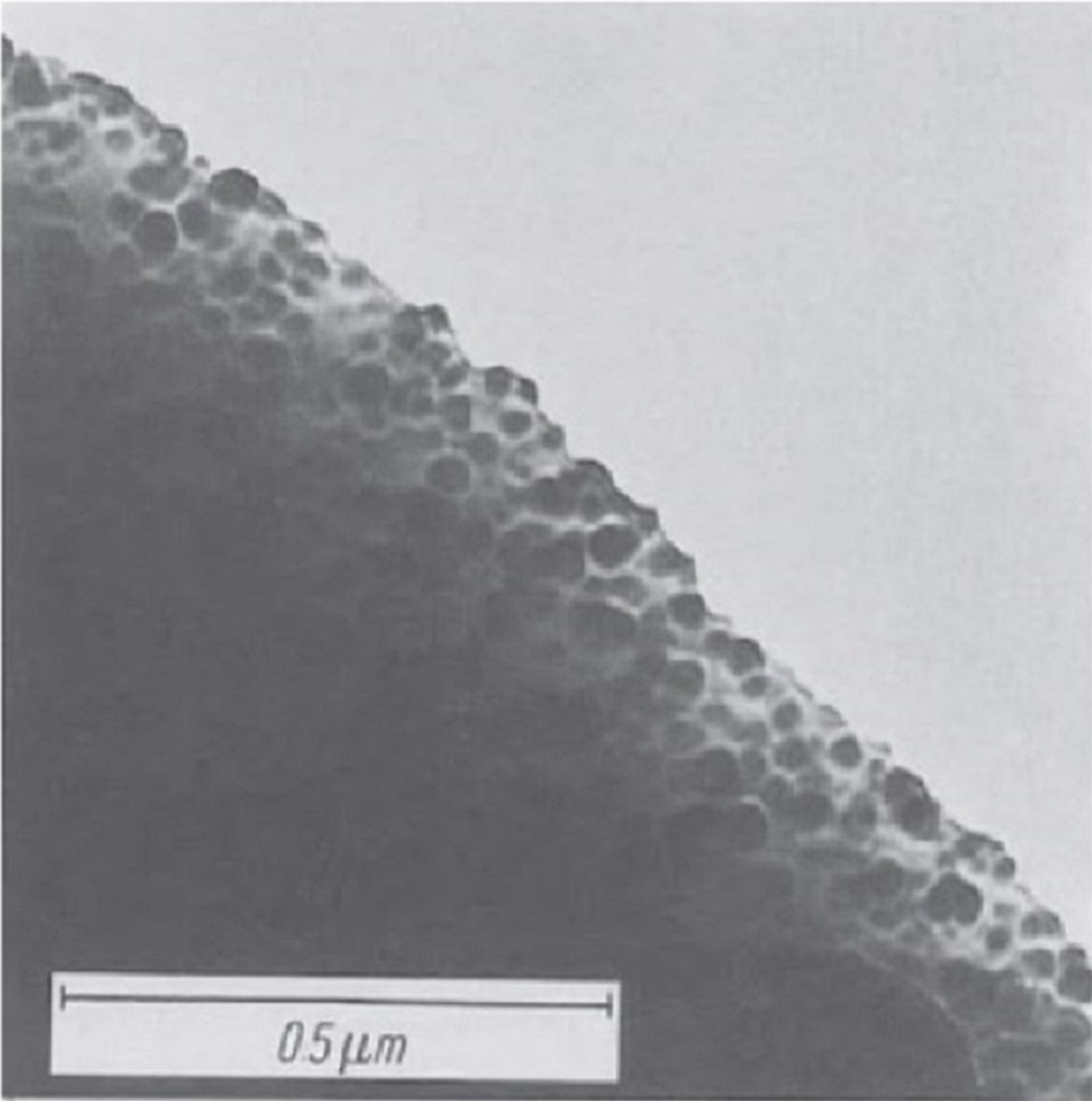}} 
 \vskip 0cm
   \subfigure[]{\includegraphics[scale=0.40] {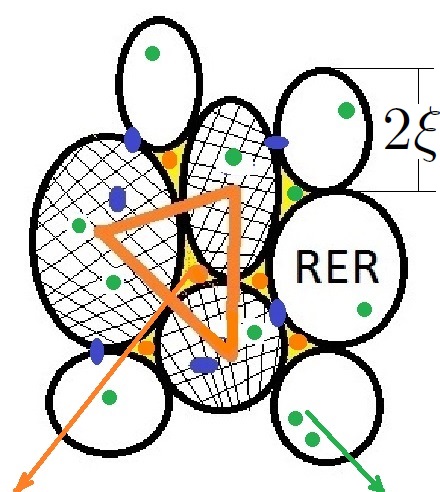} }
\vskip -0cm
   \subfigure[]{\includegraphics[scale=0.28] {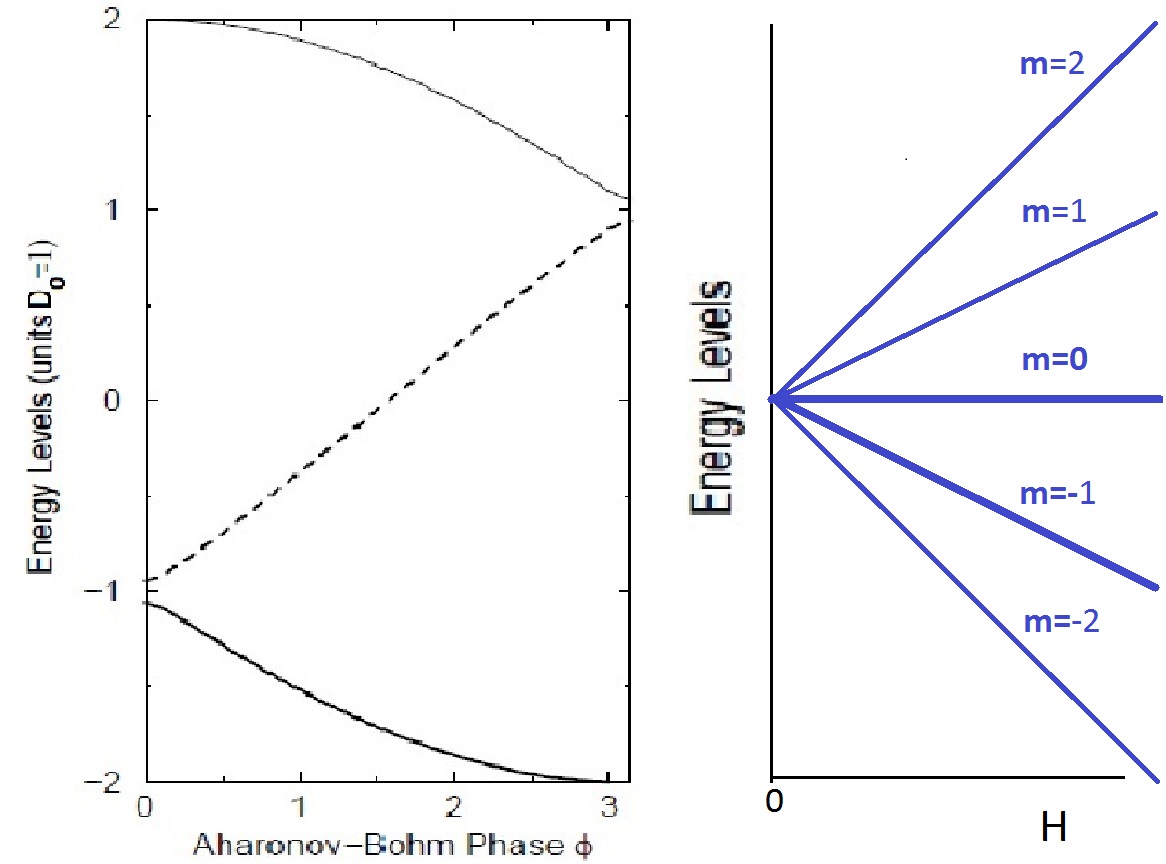} }
}
\caption{ (a) HRTEM image of a thin slice of Li$_2$O-SiO$_2$ glass (28.6 \% Li$_2$O 
molar fraction) \cite{Vog1982}; the  LiO$_2$-rich ``grains'' or ``cells'' (our own 
model's RER) are clearly visible, though not the ``fluid'' matter that comes in between.
(b) Schematic 2D representation of the structure seen in (a), with the addition of:
Fe-impurities (green); double-well tunneling defects (single-particle, blue); ATS 
(anomalous tunneling systems, collective, orange). The criss-cross pattern denotes
atomic quasi-ordering. The orange quasiparticles (on each face of a tetrahedron, in 
fact) experience a local tunneling potential having triangular topology. 2$\xi$ is the 
typical cell size (some tens of nm). Arrows point to different energy-level spectra 
below. 
(c) Left panel: Energy level spectrum of a single ATS (see \cite{Jug2004} and SI) 
as a function of the Aharonov-Bohm phase $\phi\propto H$. Right panel: A typical
energy level spectrum for a paramagnetic ion ($J_s$=2).
}
\label{model}
\end{figure}

Back to theory, 
in a nutshell the tunneling-currents  approach requires to abandon the widespread 
view of amorphous solids as continuously- and homogeneously-disordered media at 
the atomic and MR scale (the CRN model, which continues to represent a useful 
zero-order approximation). 
Instead, one admits that -- for a bulk glass -- the DH scenario characterizing the 
supercooled liquid state \cite{Edi2000} continues to be valid below $T_g$. 
Here, the ``amorphous'' solid state is made up of solid-like particle regions -- 
jammed together below $T_g$ -- and liquid-like particle regions in between
\cite{Jug2018}. The solid-like regions are also known to be better-ordered as well 
\cite{JLT2021, KAT2007,Kaw2010}, at least there is strong evidence just above 
$T_g$ \cite{Tan2018,Tan2019}, hence they are called RER (regions of enhanced 
regularity) in the glass. The particles that are more mobile -- also those proximal 
to RER walls, which are most likely charged O$^-$ dangling bonds for the 
multi-silicate glasses -- get to be squeezed together in the ``voids'' between 
the solid-like RER which themselves make up a close-packed true random solid. 
These charged, mobile particles in the ``voids'' will be highly correlated and 
subjected to complicated forces from their surroundings. A reasonable hypothesis 
\cite{Jug2004,Jug2013} is that they will tunnel coherently as effective 
quasi-particles in each highly-correlated void-contained  ``liquid drop'', the 
effective ``particles'' having large negative effective charge $Q$, $|Q|\gg|e|$ 
($e$ is the electron's charge) and heavily renormalized tunneling parameters 
\cite{Jug2004,Jug2013}. Each tetrahedral void between mostly four RER being 
composed of four faces, the quasi-particles -- one for each face 
-- will be subjected to an effective local potential having three-fold topology 
(or symmetry, for perfect RER local tetrahedral arrangement, then becoming a 
distorted three-welled tunneling potential with topological disorder in the shape, 
size and final location of the four RER). These quasi-particles are the sources 
of orbital magnetic moments (having a collective local nature) and are believed 
\cite{Bon2015} to be the sources of the observed intrinsic ``glass-magnetisation'' 
$M_{intr}^{tunn}$. More about this model of glasses in the SI and in 
\cite{Jug2018,JBK2016}.

In Fig. \ref{model}(a) we show a HRTEM (high-resolution transmission electron-
microscopy) image \cite{Vog1982} of the structure of a typical silicate glass, 
where the random closed-packed arrangement of these RER is apparent. Vogel
\cite{Vog1965} has produced many of such HRTEM pictures for network glasses,
showing that the structure here advocated for is generic for any glass 
\cite{JBK2016,Jug2018,JLT2021}. There is in fact reasonably good agreement 
between the size $2\xi$ of the RER as estimated in HRTEM 
and what is deduced from the concentration of the coherent-tunneling quasiparticles 
obtained by theory-fitting a wide spectrum of experiments on glasses in magnetic 
fields at Kelvin and sub-Kelvin temperatures \cite{Jug2018}. One gets, for BK7: 
$2\xi\simeq$ 60 nm, Duran: $2\xi\simeq$ 31 nm, BAS: $2\xi\simeq$ 32.5 nm.
Fig. \ref{model}(b) is a 2D-sketch of the proposed structure of glasses at the 
nanoscale level, with the RER containing mobile semimetal- and metal-oxide ions in 
their ``voids'' and O$^-$ dangling bonds coating their tetrahedral voids' faces. 
Their quasi-particles will be subjected to an effective three-welled tunneling potential 
\cite{Jug2018,Jug2004}.  Finally,  Fig. \ref{model}(c) [left panel] shows the ensuing 
three-level system's \cite{Boi1999} quanto-mechanical energy levels for a 
quasi-particle in the three-welled effective model \cite{Jug2004} describing the 
collective coherent-tunneling of all the $N_{tunn}$ charged O$^-$ particles sitting 
on a face of the distorted-tetrahedral void geometry. Clearly, this is a non-linear 
energy spectrum in $H$ and the idea is that this spectrum is in part responsible for 
deviations from universality in $H/T$ in the $M-M_L$ reduced magnetisation, 
Fig.s \ref{univ}(a),(b). In fact, as it turns out, the broad peak in the 
$M_{int}(T,H)$ vs. $H$ data (Fig.s \ref{intrT}(b) and \ref{otherglasses}(b)) is but a 
map of this three-level system's spectrum as a function of $\phi\propto H$ 
($\phi\equiv 2\pi{\bf B}\cdot{\bf S}_{\triangle}/\Phi_0$, where $\Phi_0=h/|Q|$ is 
the appropriate flux-quantum, $S_{\triangle}$ the surface of the flux-threaded 
triangular tunneling region and $h$ Planck's constant, see the SI and \cite{Jug2018};
${\bf B}=\mu_0 {\bf H}$ as always). 
The lower energy gap of each three-level system on a tetrahedral face opens 
non-linearly as the magnetic field increases, hence the peculiar paramagnetic 
contribution at the intermediate fields.
This theory has been employed to calculate the $M_{intr}^{tunn}(T,H)$ missing 
contribution to be added to $M_{LL}$ (Eq. (\ref{LL})) (some details in the SI, see 
also \cite{Jug2018,Bon2015,JBK2016}) and then fitted to our own present 
experimental raw data obtaining excellent fits (see SI).

The remaining phenomenon that can be accounted for is the fact that the number 
$N_{tunn}$ of coherent-tunneling charged ions on each tetrahedral face appears 
to diminish with $T$ at the lowest temperatures as was observed in fitting available 
very low-$T$ data for glasses in a magnetic field \cite{Jug2018,JBK2016,JPB2014}. 
This can be interpreted in terms of the slow growth of the RER size $2\xi$ with 
decreasing $T$ at low temperatures, but especially in terms of the broadening 
(consolidation) of the RER shape which results in the narrowing of the voids 
in-between the RER. The process can be modelled phenomenologically as 
Arrhenius-activated and an exponential $\exp\{E_0/k_BT\}$ is then involved 
(see SI). The fact that typical values of $E_0$ from the best-fit of typical 
SQUID-run data vs. $T$ are never larger than 0.5 K (0.043 meV) is an indication 
that adsorption of positively-charged metal- and semimetal-oxide ions from the 
RER-voids-contained ``melt'' onto the tetrahedral walls is enhanced by the 
presence of a dense covering of existing O$^-$ dangling bonds. Typical activation 
energy values for Si$^{4+}$, B$^{3+}$, Al$^{3+}$, K$^{+}$, Na$^{+}$ etc. 
adsorbing to a clean SiO$_2$ wall are in fact in the meV to fraction of eV range 
\cite{Mem2001}. The covering of an RER-void interface by a O$^-$ dangling-bond 
layer gives rise to a highly-correlated system of tunneling charged particles in a 
magnetic field. As the RER-size grows at the expense of the residual melt in each 
void, the area becoming available to O$^-$ species diminishes and $N_{tunn}$ drops
exponentially with decreasing temperatures (zipping-up of the RER-RER interface, see 
SI). We believe it is this process -- crucial to the onset of the cryogenic (sub-Kelvin) 
properties of glasses -- that is responsible for the ubiquitous broad peak of the 
$M_{intr}$ as a function of $1/T$ at the lowest temperatures. The 
position of this peak depends almost only on $E_0$ and for this reason the peak 
basically does not shift with changing $H$, contrary to the peak in $M_{intr}(T,H)$ 
vs. $H$ at fixed $T$, which moves to higher values of $H$ as $T$ decreases. As is 
indeed observed: the agreement between theory and experiments is very satisfactory.
Finally, the new -- unexpected -- observation in our SQUID-measurements is that 
there are possible further oscillations in $M_{intr}$ at the lowest temperatures 
investigated (see Fig.s \ref{intrH}(a),(b) and \ref{otherglasses}(a)) beside those 
already mentioned at intermediate temperatures (see Fig. \ref{oscill}). A hint of 
why these come about can be found in the SI (crossover from 2D to 3D 
correlated/coherent tunneling regions).

Regarding the interpretation of the data, we remark that it might be argued that 
at high $H/T$ the data might be consistent with a constant value of $M_{intr}$. 
However: 1) We know of no mechanism producing a constant $M_0$ for $H$ 
above a finite threshold ($H_0$) and/or for $1/T$ above a $1/T_0$; 2) The
offered theoretical interpretation gives good fits with $M_{intr}$ data presenting 
a broad peak for high $H/T$ and then slowly decreasing (see the SI), but there 
might be further oscillations due to new collective phenomena.   

\vfill
{\bf F. Conclusions.} We have reported that an exotic new form of (para-)magnetism 
can be revealed in the measurement of the SQUID-magnetisation in glasses and is 
solely due to the nanoscale structure of the vitreous state itself. Though very small 
(for comparison: Fe (218 emu/g), Ni (55 emu/g) at 300 K (spontaneous bulk
magnetisation)) due to low concentrations and the fact that the coherent tunneling 
currents' magnetic moments tend to cancel each other out in each tetrahedral void, 
the $M_{intr}(T,H)$ intrinsic magnetisation is very interesting in its unusual $T$- 
and $H$-dependence and represents a new way to look -- experimentally -- at 
glass structure and other physical properties of glasses (of all types, in fact). An 
important consequence of our findings when interpreted through a nanoscopic-scale 
cellular theory for the vitreous state's organisation is the existence of a novel 
type of tunneling states that residing in the voids between compact nanoscopic 
solid-like regions appear to survive till room temperature because of the 
{\em protection} provided by the surrounding RER. We have proposed a new 
research tool for the study of the physics of glasses.
Studying the weak but intrinsic glass magnetism in different glass types should be 
a very interesting new way to look at the glass ``transition'' and of modelling the 
physical properties of all glasses in an effective and efficient new way. 
The {\em protected} new tunneling systems \cite{Jug2004}, the existence of which 
\cite{Boi1999} is confirmed by this work, promise to be very interesting for applications 
to quantum information technology due to their own inner natural coherence.
 
\vskip 1 cm
{\bf Materials and methods.} Extreme care was taken not to accidentally contaminate 
the samples, which were first SQUID- and then MS-analysed (BK7 only). A BK7
prism of ca. 3 cm length and 5 mm through size was purchased from Edmund Optics,
Inc., Barrington NJ (USA) and chips were obtained by means of a ceramic knife and 
rubber hammer blows. Duran samples were obtained by shattering (with the rubber 
hammer) a brand new Schott chemistry beaker and selecting appropriate shards for 
measurement. The two BAS samples where obtained by firing the Heraeus 
IP-211-clear sol-gel paste in small Pt-foil vessels and then carefully peeling off the foil 
after obtaining perlaceous, glassy consistency.

The dc magnetisation $M$ of the studied glasses was measured by means of a 
Quantum Design MPMS-XL$7$ magnetometer based on a SQUID (Superconducting 
QUantum Interference Device). Cotton thread was used to suspend each sample in 
the center of a plastic cannula which was then inserted in the magnetometer. It 
was always ensured that the linear dimensions of the samples along the cannula's 
axis were below $\sim 5$ mm in order to guarantee a reliable instrumental estimate 
of the magnetisation value from the voltage measured along the detection coils. 
Although the studied samples had different shapes, the weak measured absolute 
values of $M$ make it safe to neglect the effects of geometric demagnetisation. 
The plastics of the cannula is also a type of (polymeric) glass, however the amount 
of material involved (thin wall) is reputed to be small in comparison to the samples'. 
The large size of the polymer's RER, thus the tiny concentration of the polymer 
glass' intrinsic magnetic moments, is also thought to ensure a negligible contribution. 

The investigated ranges for temperature and magnetic field were 
$2$ K $\leq T \leq 320$ K and $0$ kOe $\leq H \leq 65$ kOe, respectively. The 
temperature scans were performed at constant magnetic field always while warming 
the sample after a field-cooling procedure from room temperature. Especially at low 
temperatures ($T \leq 60$ K), where the temperature dependence of $M_{intr}$ is 
more pronounced, slow warming rates $\sim 0.3$ K/min were used in order to reduce 
the possible effect of systematic errors in the temperature reading. The magnetic field 
scans were performed at constant temperature after a zero-field-cooling procedure.

ICP-MS (simply MS, in this paper) determinations were performed with a 
Thermo-Fisher ICAP-Q spectrometer. For BK7 and Duran weighted splinters of glass 
(less than 100 mg) were dissolved with ultra-pure acids (1.5 ml of HF + 0.5 ml of HCl) 
over a hot plate in mild boiling conditions for 90 mins. For each sample residual HF 
is moved away with two subsequent additions of HNO$_3$ (1 ml each) under boiling 
conditions. Samples are then MS-analyzed after proper dilution.

Heraeus-IP-211-clear glass (BAS) was dissolved in harder conditions. Typically, a 
(ca.) 30mg specimen is treated with 1ml of HF plus 0.6 ml of HCL plus 0.2 ml of 
HNO$_3$ in a sealed vessel of a Milestone Ethos One microwave oven. 450 min at 
180°C were necessary to reach complete dissolution, due to the great difficulty in 
dissolving aluminosilicate clusters. Even in this case residual HF is moved away with 
two subsequent additions of HNO$_3$ (1 ml each) under boiling conditions. 

\vfill
{\bf Acknowledgements. } GJ acknowledges support by INFN-Pavia through Iniziativa 
Specifica GEO-SYM, also useful conversations with K. Bevan about silicate chemistry 
and with R. Santoro for hints on data-fitting. The Authors are very grateful to Giacomo 
Prando and Pietro Carretta of the Universit\`a di Pavia, Italy, for many discussions on 
this project, for carrying out the SQUID-magnetisation measurements for us and for 
generously providing the raw data to the PI.

\vfill
{\bf Author Contributions. } The Principal Investigator (GJ) conceived the project, 
chose the systems to be investigated, carried out the SQUID-data analysis and wrote
the manuscript. SR made the BAS-glass samples and carried out the ICP-MS analyses 
of all the samples.

\vfill
{\bf Data availability.} The data that support the plots and tables within this paper 
and other findings of this study are available from the corresponding author upon 
reasonable request.

\vfill
{\bf Additional information.} Supplementary Information (SI) is available for this paper.

\vfill
{\bf Conflicting interests.} The Authors declare no competing financial interests.



\bibliographystyle{unsrt}


\vfill
\end{document}